\documentclass[10pt,aps,prd,amsmath,preprintnumbers,showpacs,nofootinbib,superscriptaddress,notitlepage,twocolumn, amssymb, usenatbib]{revtex4-2}

\usepackage{bm}

\usepackage[T1]{fontenc}
\usepackage{ae,aecompl}

\usepackage{graphicx}	
\usepackage{amsmath}	
\usepackage{amssymb}	
\usepackage{comment}    
\usepackage{siunitx}
\usepackage{url}

\makeatother
\newcommand{\citelink}[1]{\hyperlink{cite.\{biblio.bib}}

\newcommand{\vect}[1]{\boldsymbol{\mathbf{#1}}}

\usepackage{todonotes} 

\begin{document}

\preprint{APS/123-QED}

\title[]{A Pulsar-Based Map of Galactic Acceleration} 

\author{Abigail Moran}
\email{abigail.moran@columbia.edu}
\affiliation{Department of Astronomy, Columbia University,  New York, NY 10027, USA}
\affiliation{Department of Physics, University of Connecticut, 196 Auditorium Road, U-3046, Storrs, CT 06269-3046, USA}

\author{Chiara M. F. Mingarelli}
\affiliation{Department of Physics, Yale University, New Haven, CT 06520, USA}
\affiliation{Department of Physics, University of Connecticut, 196 Auditorium Road, U-3046, Storrs, CT 06269-3046, USA}
\affiliation{Center for Computational Astrophysics, Flatiron Institute, 162 Fifth Ave, New York, NY 10010, USA}

\author{Ken Van Tilburg}
\affiliation{Center for Computational Astrophysics, Flatiron Institute, 162 Fifth Ave, New York, NY 10010, USA}
\affiliation{Center for Cosmology and Particle Physics, Department of Physics,New York University, New York, NY 10003, USA}

\author{Deborah Good}
\affiliation{Department of Physics and Astronomy, University of Montana, 32 Campus Drive, Missoula, MT, 59812 USA}
\affiliation{Department of Physics, University of Connecticut, 196 Auditorium Road, U-3046, Storrs, CT 06269-3046, USA}
\affiliation{Center for Computational Astrophysics, Flatiron Institute, 162 Fifth Ave, New York, NY 10010, USA}

\date{\today}

\begin{abstract}
\noindent Binary pulsars can be used to probe Galactic potential gradients through calculating their line-of-sight accelerations. We present the first data release of direct line-of-sight acceleration measurements for {29} binary pulsars. We validate these data with a local acceleration model, and compare our results to {those from earlier works}. We find evidence for an acceleration gradient in agreement with these values, with our results indicating a local disk density of ${\rho_\mathrm{d} = 0.040_{-0.020}^{+0.020}} \ \mathrm{M_\odot}\mathrm{pc}^{-3}$. We also find evidence for unmodeled noise of unknown origin in our data set.

\end{abstract}

\maketitle

\section{Introduction}
Characterizing the distribution of matter in  the Milky Way (MW) is crucial for determining the microphysics of dark matter and the dynamical history of our Galaxy. The MW's mass distribution has been partially constrained by its circular rotation curve~\cite{eilers2019circular, benito2021uncertainties}. These measurements, in conjunction with baryonic mass models, can determine the dark matter distribution~\cite{baryonic}. Several analyses~\cite{kuijken1989mass,holmberg2000local,bovy2012local,widmark2019measuring, Mckee2015} of velocity-distance data have estimated the midplane density assuming the MW is in dynamic equilibrium, an assumption now known not to hold~\cite{bennett2019vertical, antoja2018dynamically,naik2022charting, Haines}. 

Stellar velocities from future spectroscopic surveys may serve as a sensitive, direct probe of Galactic acceleration~\cite{Quercellini, silverwood2019stellar,chakrabarti2020toward, ravi2019probing, NG15data}. Other model-independent dark matter measurements with transverse angular accelerations~\citep{buschmann2021galactic} are possible using \textit{Gaia} astrometry~\cite{Gaiacollab}. Even the fine-grained substructure in the MW may be detectable with pulsar timing arrays (PTAs)~\citep{dror2019pulsar, ramani2020observability}, through gravitational lensing of stellar motions~\cite{van2018halometry},{ and via observations of eclipsing binaries~\cite{eclipsingbinaries}}.

The ultra-stable millisecond pulsars (MSPs) used in PTA experiments are normally used to search for gravitational waves (GWs)~\citep{sazhin,det79,HD83,Ming2022,GWDetection, NG15data} and to test general relativity~\citep{GenRelTest, GRwRadio, Scalarization, Strongfieldtest}. 
Most of these pulsars have binary companions~\cite{Lorimer05}, and secular changes in their orbital periods $P_\mathrm{b}$ are measurable for a select few of these. There are also a few precise standard binary pulsars {(not MSPs)} which have measured orbital period derivatives, $\dot{P}_\mathrm{b}$.

Measurements of $\dot{P}_\mathrm{b}$ have been used to estimate pulsar distances through the Shklovskii effect~\cite{Shk70,Verbiest} assuming a fixed Galactic acceleration model. One can also evaluate the predictions of modified gravity theories by studying $\dot{P}_\mathrm{b}$ \citep{Pb_GW}, as well as search for GWs with frequencies below $10^{-9}\, \mathrm{Hz}$~\cite{DeRocco:2022irl,DeRocco:2023qae}. Recently, there has been interest to use well-timed binary pulsars to directly measure line-of-sight accelerations~\cite{Philips, Chakrabarti, Heflin}. This acceleration is a direct probe of the relative gradient of the MW's gravitational potential at the pulsars' locations.

Previously, Ref.~\cite{Heflin} reported this acceleration for PSR J1713+0747. Refs.~\cite{Philips,Chakrabarti} also measured Galactic accelerations using 13 and 14 binary pulsars, respectively. However, Refs.~\cite{Philips,Chakrabarti} did not directly report individual accelerations nor their associated uncertainties, which we do here for the first time.

Our main result is an open data release of directly measured Galactic accelerations, uncertainties, and other supplementary observables for {29} pulsars {-- the largest catalog of these accelerations to date}. We also perform parameter estimation for the local acceleration gradients and density of the Galactic disk's midplane {as in \cite{Chakrabarti}}, and include for the first {time a model} that includes spurious acceleration noise {to accommodate the existence of several outliers}.

The paper is laid out as follows: in Sec.~\ref{sec:acc}, we describe how to directly infer line-of-sight acceleration from the pulsar data, and a simple model for Galactic acceleration. In Sec.~\ref{sec:pulsars}, we detail our pulsar selection criteria. 
We benchmark our catalog in Sec.~\ref{sec:results} with a Markov Chain Monte Carlo (MCMC) analysis to estimate the parameters of this Galactic acceleration model. A summary of our catalog can be found in Tab.~\ref{tab:bigtab} of App.~\ref{app:catalog}, and an accompanying $\texttt{.csv}$ file is available with this paper~\cite{Supplemental}. 

\section{Galactic Acceleration}\label{sec:acc}

The observed time derivative of a binary pulsar's orbit, $\dot{P}_\mathrm{b}^\text{Obs}$, is a combination of several factors~\cite{Bell}:
\begin{equation}\label{eq:pbdot}
\dot{P}_\mathrm{b}^\text{Obs} = \dot{P}_\mathrm{b}^\text{Kin} + \dot{P}_\mathrm{b}^\text{GW} + \dot{P}_\mathrm{b}^\text{Gal} + \dot{P}_\mathrm{b}^\text{n} ,
\end{equation}
where $\dot{P}_\mathrm{b}^\text{Kin}$ arises from kinematic effects \citep{Shk70}, $\dot{P}_\mathrm{b}^\text{GW}$ from GW emission \citep{GW}, and $\dot{P}_\mathrm{b}^\text{Gal}$ from Galactic acceleration. {Due to the presence of several large outliers (see Section \ref{sec:results}), }we introduce $\dot{P}_\mathrm{b}^\text{n}$, a noise term encapsulating hidden nuisance effects on the binary period due to undetected companions~\cite{Nitu, pulsar_planets, 3rd_body_constraints, J0210}, accretion, or other unknowns (e.g. baryon loss \cite{BNV1, BNV2}).  

The kinematic term is also called the Shklovskii effect~\cite{Shk70}:
\begin{equation}\label{eq:shklovskii}
    \dot{P}^\text{Kin}_\mathrm{b} = \mu^2 r \frac{P_\mathrm{b}}{c} \, ,
\end{equation}
where $\mu$ is the proper motion of the pulsar, $r$ is its distance, and $c$ is the speed of light. 
The GW emission term~\cite{Shapiro} is:
\begin{equation} \label{eq:GW}
        \dot{P}^\text{GW}_\mathrm{b} = - \frac{192 \pi G^{\frac{5}{3}}}{5c^{5}} F(e) \left(\frac{P_\mathrm{b}}{2\pi}\right)^{-\frac{5}{3}} 
        \frac{M_\mathrm{p} m_\mathrm{c} }{\left(M_\mathrm{p} + m_\mathrm{c}\right)^{\frac{1}{3}}} \, ,
\end{equation}
where $e$ is the eccentricity of the binary orbit, $F(e) = (1-e^2)^{-\frac{7}{2}}
        (1+\frac{73e^2}{24}+\frac{37e^4}{96})$ which is unity for circular orbits, and $M_\mathrm{p}$ and $m_\mathrm{c}$ are the masses of the pulsar and companion, respectively. 
{For PSR J1455-3330, which has no pulsar or companion mass measurement, we approximate $\dot{P}^\text{GW}_\mathrm{b} = 0$. This pulsar is in a 76 day, nearly circular orbit (e $\sim 10^{-4}$), and this term is thus expected to be subdominant to $\dot{P}^\mathrm{Kin}_\mathrm{b}$ \cite{EPTADR2}.}

We can compute $\dot{P}_\mathrm{b}^\text{Gal}$ {as in Ref.~\cite{Chakrabarti}} by subtracting the kinematic and GW terms from $\dot{P}_\mathrm{b}^\text{Obs}$ and marginalizing over the noise contribution $\dot{P}_\mathrm{b}^\text{n}$ in Eq.~\ref{eq:pbdot}. We can furthermore write the differential Galactic line-of-sight acceleration as
\begin{equation}
    \vect{a}_\text{Gal} \cdot \hat{\vect{r}} = c  \frac{\dot{P}_\mathrm{b}^\text{Gal}}{P_\mathrm{b}} = -\bigg[ \vect{\nabla}\Phi(\vect{r}_\text{b}) -  \vect{\nabla}\Phi(\vect{r}_\odot) \bigg]  \cdot \hat{\vect{r}}, \label{eq:a_est}
\end{equation}
where $\hat{\vect{r}} \equiv \vect{r} / |\vect{r}|$,  the distance to the pulsar is $\vect{r} = \vect{r}_\text{b} - \vect{r}_\odot$, 
and $\Phi$ is the gravitational potential. The difference in gradient of the gravitational potential between the Earth ($\vect{r}_\odot$) and the binary pulsar ($\vect{r}_\text{b}$) can be directly measured with $P_\mathrm{b}$ and $\dot{P}_\mathrm{b}^\text{Gal}$.

We estimate the parameters of the following model for the Galactic acceleration:
\begin{align}
    \vect{a}_\text{Gal}(\vect{r}) = a'_x \, \hat{\vect{x}} (\hat{\vect{x}} \cdot \vect{r})  + a'_z \, \hat{\vect{z}} (\hat{\vect{z}} \cdot \vect{r}),
    \label{eq:a_gal_model}
\end{align}
where $\hat{\vect{x}}$ and $\hat{\vect{z}}$ are unit vectors in galactocentric coordinates, pointing away from the Galactic Center and towards the Galactic North Pole (perpendicular to the disk), respectively. The fit parameters in Eq.~\ref{eq:a_gal_model} are the Taylor expansion coefficients of the local acceleration field, with $a'_x$ directly related to the slope of the MW's rotation curve, and the vertical acceleration gradient $a'_z$ (primarily) to the density of the Galactic disk $\rho_\mathrm{d}$ through Poisson's equation {for Newtonian gravity: $\vect{\nabla} \cdot \vect{a} = - 4 \pi G \rho$. Locally, we can model the galactic disk as a slab of nearly uniform density $\rho_\mathrm{d}$ and essentially infinite extent, which translates to a vertical acceleration gradient:}
\begin{align}
    a'_z \simeq 4\pi G \rho_\mathrm{d} \, .
    \label{eq:rho}
\end{align}
This local {Taylor} expansion should be a good approximation since the pulsars in our sample populate the local Galactic neighborhood and are predominantly inside the galactic disk (more in Sec.~\ref{sec:pulsars}).

\section{Pulsar Catalog Construction}\label{sec:pulsars}
We search the Australia Telescope National Facility's (ATNF) pulsar catalog~\cite{ATNF} for all pulsars with $P_\mathrm{b}$ and $\dot{P}_\mathrm{b}^\mathrm{Obs}$ measurements.
We select pulsars with measured proper motion, and distance based on astrometric parallax or timing parallax, necessary to calculate the Shklovskii effect in Eq.~\ref{eq:shklovskii}. 
If a pulsar is missing a parallax measurement, we try to obtain the distance to the companion (and thus the binary) using techniques from Ref.~\cite{Moran}.
We use the combined distances for J0437-4715 and J1012+5307 from Ref.~\cite{Moran}. We also identify {\it Gaia} DR3 3273288485744249344 as PSR J0348+0432's companion, {with a detection significance of $4.5\sigma$. This object has a parallax of $-0.035 \pm 0.784$ mas, which translates to a distance of $1600 \pm 900$ pc using the distance prior from Ref.~\cite{bailer-jones2021}. Since the parallax is negative, the resulting distance is heavily dependent on the choice of distance prior. However, this is the first parallax measurement to the system ever, and future \textit{Gaia} data releases may improve this measurement.}

We exclude pulsars in globular clusters since these have accelerations that are not expected to conform to our simple model in Eq.~\ref{eq:a_gal_model}. We further exclude pulsars which are accreting mass from their companions (so-called ``spider pulsars'') since this can cause the binary period to decrease~\cite{spinup}.
This selection returns {29} pulsars, see Figs.~\ref{fig:accmap} and \ref{fig:zmap} in the App.~\ref{app:catalog} for their distribution in the Galactic plane.

{Recently Ref. \cite{Donlon24} used a slightly modified version of our catalog to measure galactic acceleration and constrain the parameters of several models of gravitational potential. They omitted five pulsars from our original catalog, citing concerns over possible mass transfer between the pulsars and their companions.
To ensure a robust catalog, we keep all pulsars which have not been definitively identified as undergoing mass transfer or as spider systems. We explore how removing potential (though unconfirmed) pulsars undergoing mass transfer sources would affect our analysis in Sec.~\ref{app:mod_catalog}.} 
    
\section{Results}\label{sec:results}

We directly compute the line-of-sight acceleration for the {29} pulsars in our catalog---the largest of its kind, via Eq.~\ref{eq:pbdot} and the first equality in Eq.~\ref{eq:a_est} to obtain:
\begin{align}
a_i = c P_{\mathrm{b},i}^{-1}
\left[\dot{P}_{\mathrm{b},i}^\text{Obs} - \dot{P}_{\mathrm{b},i}^\text{Kin} - \dot{P}_{\mathrm{b},i}^\text{GW} \right] \, . \label{eq:a_est2}
\end{align}

\begin{figure}
\includegraphics[trim = 15 10 25 35, clip, width=1\linewidth]{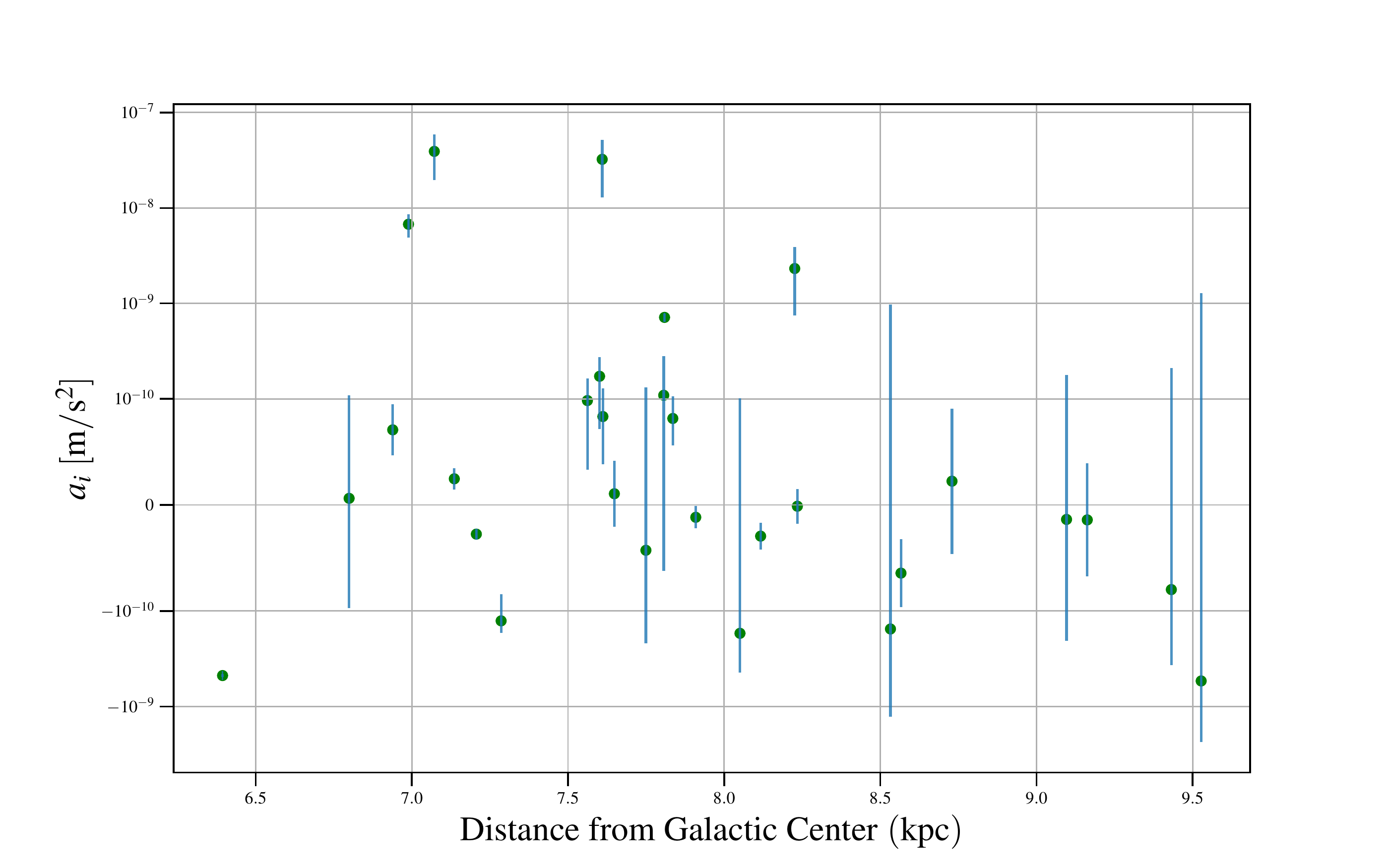}
\caption{Line-of-sight acceleration of each binary pulsar as a function of distance from the Galactic Center. Error bars include uncertainties propagated from measurements of proper motion, distance, and timing parameters only, and do not include the additional acceleration noise parameter, $\dot{P}_\mathrm{b}^\text{n}$.}
\label{fig:accs}
\end{figure}

 The observational uncertainties on $a_i$,  denoted by $\sigma_{a,i}$, are the quadrature sums of the uncertainties for each of the terms in Eq.~\ref{eq:a_est2}. 
In Fig.~\ref{fig:accs}, we show $a_i$ and $\sigma_{a,i}$ as a function of distance from the Galactic Center. Figs.~\ref{fig:accmap} and \ref{fig:zmap} in App.~\ref{app:catalog} show maps of these data.

To assess the quality and sensitivity of our pulsar acceleration catalog, we perform an MCMC analysis of the parameters of the simple local acceleration model of Eq.~\ref{eq:a_gal_model}.
We find that the sum of kinematic, GW, and Galactic acceleration contributions does not adequately explain the data: a small (but larger than expected) subset of binary period derivatives are {outliers, with accelerations larger than expected in physically plausible large-scale models of the MW,} (and are several standard deviations from the best fit predictions of Eq.~\ref{eq:a_gal_model}). {Under the assumption of Gaussian errors, these pulsars skew the fit parameters of Eq.~\ref{eq:a_gal_model} as well as any goodness of fit calculations.} 

{Motivated by the presence of these large outliers,} we postulate that, with a probability $p_\mathrm{n}$, any binary pulsar is subject to an additional, unknown acceleration noise $\vect{a}_\mathrm{n} = a_\mathrm{n} \hat{\vect{R}}_\mathrm{n}$ as a nuisance parameter in a random direction $\hat{\vect{R}}_\mathrm{n}$.
The probability density function (PDF) for the angle $\theta_\mathrm{n} \equiv \arccos(\hat{\vect{a}}_\mathrm{n} \cdot \hat{\vect{r}}_i)$, where $0\leq \theta_\mathrm{n} \leq \pi$,  is thus $\sin(\theta_\mathrm{n})/2$. The magnitude of the acceleration is taken to have the normalized PDF $C a_\text{n}^{-\zeta}$ with normalization coefficient $C = (\zeta-1) / [(a_\mathrm{n}^\mathrm{min})^{-\zeta+1}-(a_\mathrm{n}^\mathrm{max})^{-\zeta+1}]$. 
The minimum noise acceleration is $a_\mathrm{n}^\text{min} \approx 1.77 \times 10^{-12} \, \mathrm{m}\,\mathrm{s}^{-2}$ (less than the smallest inferred acceleration error), corresponding to the constant acceleration from a gravitational perturber of mass $1\,M_\oplus$ and semimajor axis $100\,\mathrm{AU}$ in a circumbinary orbit.
The maximum noise acceleration is $a_\mathrm{n}^\text{max} = 2.94 \times 10^{-6} \, \mathrm{m}\,\mathrm{s}^{-2}$, corresponding to a third body with $0.2\, M_\odot$ mass and $20\,\mathrm{AU}$ semimajor axis.
We are not claiming to have detected triple systems, but this is a physical and plausible example of a model which can generate additional acceleration noise. 

The addition of this nuisance acceleration term $\vect{a}_\mathrm{n}$ leads to the following likelihood function $\mathcal{L}$ for the observed acceleration:
\begin{alignat}{2}
&\mathcal{L}\left(\lbrace a_i \rbrace | a_x',a_z',p_\mathrm{n},\zeta \right) = \int_0^\pi \mathrm{d} \theta_\mathrm{n} \, \frac{\sin \theta_\mathrm{n}}{2} \int_{a_\mathrm{n}^\mathrm{min}}^{a_\mathrm{n}^\mathrm{max}} \mathrm{d}a_\mathrm{n}\, C a_\mathrm{n}^{-\zeta} \nonumber \\
& \prod_{i=1}^{N} \bigg[  (1-p_\mathrm{n}) f(\Delta a_i,\sigma_{a,i}) +p_\mathrm{n} f(\Delta a_i-a_\mathrm{n} \cos \theta_\mathrm{n},\sigma_{a,i}) \bigg]. \label{eq:likelihood}
\end{alignat}
The likelihood function takes as input the line-of-sight acceleration estimates $\lbrace a_i \rbrace$ for each pulsar $i = 1,\dots,N$. The function $f(a,\sigma)$ is a normal distribution in $a$ with standard deviation $\sigma$. In Eq.~\ref{eq:likelihood}, we have abbreviated $\Delta a_i \equiv a_i - a_{\mathrm{Gal},i}$ and $a_{\mathrm{Gal},i} \equiv \vect{a}_\mathrm{Gal}(\vect{r}_i) \cdot \hat{\vect{r}}_i$ from Eq.~\ref{eq:a_gal_model} defined in terms of $a_x'$ and $a_z'$.

\begin{figure}[t]
 \centering
     \includegraphics[width=.48\textwidth]{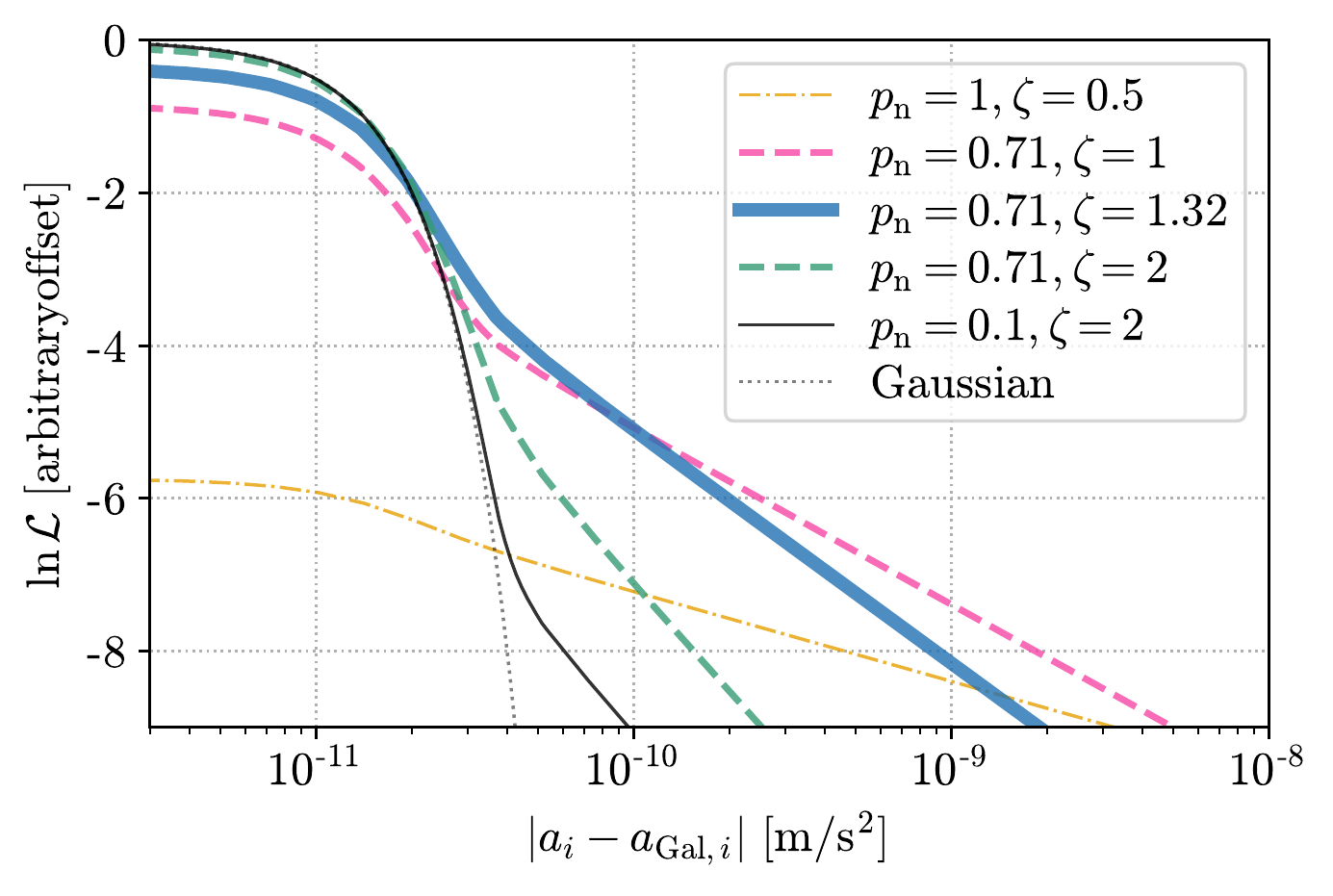}
     \vspace{-2\baselineskip}
     \caption{Log likelihood $\ln \mathcal{L}$ from Eq.~\ref{eq:likelihood} (with a fixed global offset) for a single pulsar with (observational) acceleration uncertainty of $\sigma_{a,i} = 10^{-11} \, \mathrm{m/s^2}$, as function of $|a_i - a_{\mathrm{Gal},i}|$, the difference from the acceleration model prediction. The blue curve represents the best-fit model from Fig.~\ref{fig:zeta_corner}, which is essentially a Gaussian likelihood for small acceleration deviations, but one with fatter tails whose height and slope are controlled by the nuisance parameters $p_\mathrm{n}$ (the probability of additional noise) and $\zeta$ (the spectral index). We also plot the log likelihoods for other reference values of $p_\mathrm{n}$ and $\zeta$ \textbf{for a set $a_\mathrm{n}^\mathrm{max}$ and $a_\mathrm{n}^\mathrm{min}$}, as well as that of a Gaussian distribution (dotted curve).}
     \label{fig:nuisance_model}
 \end{figure}

{We emphasize that this nuisance acceleration parameterized by $p_\mathrm{n}$ and $\zeta$ may be due to accretion, other acceleration (from e.g.~a third body), or any other unknown factor in the pulsar system.}
By introducing these parameters we produce a model which accounts for nuisance acceleration measurements without removing apparent outliers {or making any assumptions about the cause of the these nuisance measurements. $p_\mathrm{n}$ and $\zeta$ parameterize modifications to a Gaussian distribution, and do not specify the particular source of any detected noise.} This modified distribution better reflects the spread in the acceleration data, due to its heavier tail than that of a Gaussian distribution (see e.g.~Fig.~\ref{fig:nuisance_model}).

In Fig.~\ref{fig:nuisance_model}, we show the resulting likelihood function $\mathcal{L}$ for a \emph{single} pulsar with uncertainty $\sigma_a = 10^{-11} \, \mathrm{m/s^2}$ as a function of $|a_i - a_{\mathrm{Gal}_i}|$. In the limit of $p_\mathrm{n} \to 0$ or $\zeta \to \infty$, it reduces to a normal distribution. For non-zero $p_\mathrm{n}$ and finite $\zeta$, the distribution has fatter tails with height proportional to $p_\mathrm{n}$ and logarithmic slope controlled by $\zeta$. As long as $\zeta > 1$, the variance of the distribution is essentially independent from the choice of $a_\mathrm{n}^\mathrm{max}$, and will not dramatically alter the typical posterior widths (only the tails). 

\begin{table}[b!]
    \centering
    \renewcommand{\arraystretch}{1.5} 
    \begin{tabular}{l S[table-format=1.2]@{\hspace{0.7em}} S[table-format=1.2]@{\hspace{0.7em}} S[table-format=2.1]@{\hspace{0.7em}}}
    \hline 
    \hline
    {Parameter} & {$a'_x$} & {$a_z'$} & {$\rho_\mathrm{d}$} \\
    {Units} & {$10^{-10}\,\mathrm{\frac{m}{s^2 \, kpc}}$}  & {$10^{-10}\,\mathrm{\frac{m}{s^2 \, kpc}}$} & {$10^{-2}\,\frac{M_\odot}{\mathrm{pc}^{3}}$} \\
    \hline
    {This Work} & {${0.39_{-0.15}^{+0.15}}$} & {${-0.70^{+0.36}_{-0.35}}$} & {${4.0_{-2.0}^{+2.0}}$} \\
    {Ref.~\cite{Kipper_model}} & {$0.16^{+0.54}_{-0.50}$} & {$-0.98^{+0.54}_{-0.55}$} & {$5.6^{+3.1}_{-3.1}$} \\
    {Ref.~\cite{holmberg2000local}} & {---} & {$-1.78^{+0.18}_{-0.18}$} & {$10.2^{+1.0}_{-1.0}$} \\
    {Ref.~\cite{Creze}} & {---} & {$-1.3^{+0.3}_{-0.3}$} & {$7.6^{+1.5}_{-1.5}$} \\
    {Ref.~\cite{Chakrabarti}} & ${0.35^{+0.12}_{-0.16}}$ & {$-1.52_{-0.37}^{+0.83}$} & {$8^{+5}_{-2}$} \\
    \hline
    \hline
    \end{tabular}
    \caption{
    Best-fit parameters for the Galactic acceleration gradients $a_x'$ and $a_z'$ in the radial and vertical directions (cfr.~Eq.~\ref{eq:a_gal_model}), respectively, and corresponding density $\rho_\mathrm{d}$ in the midplane of the MW disk via Eq.~\ref{eq:rho}. {For Refs.~\cite{holmberg2000local, Creze}, we calculate $a_z'$ from the reported value of $\rho_\mathrm{d}$ and Eq.~\ref{eq:rho}.}
    }
    \label{tab:mod_params}
\end{table}

\begin{figure}[b]
\vspace{-1\baselineskip}
\includegraphics[width=1\linewidth]{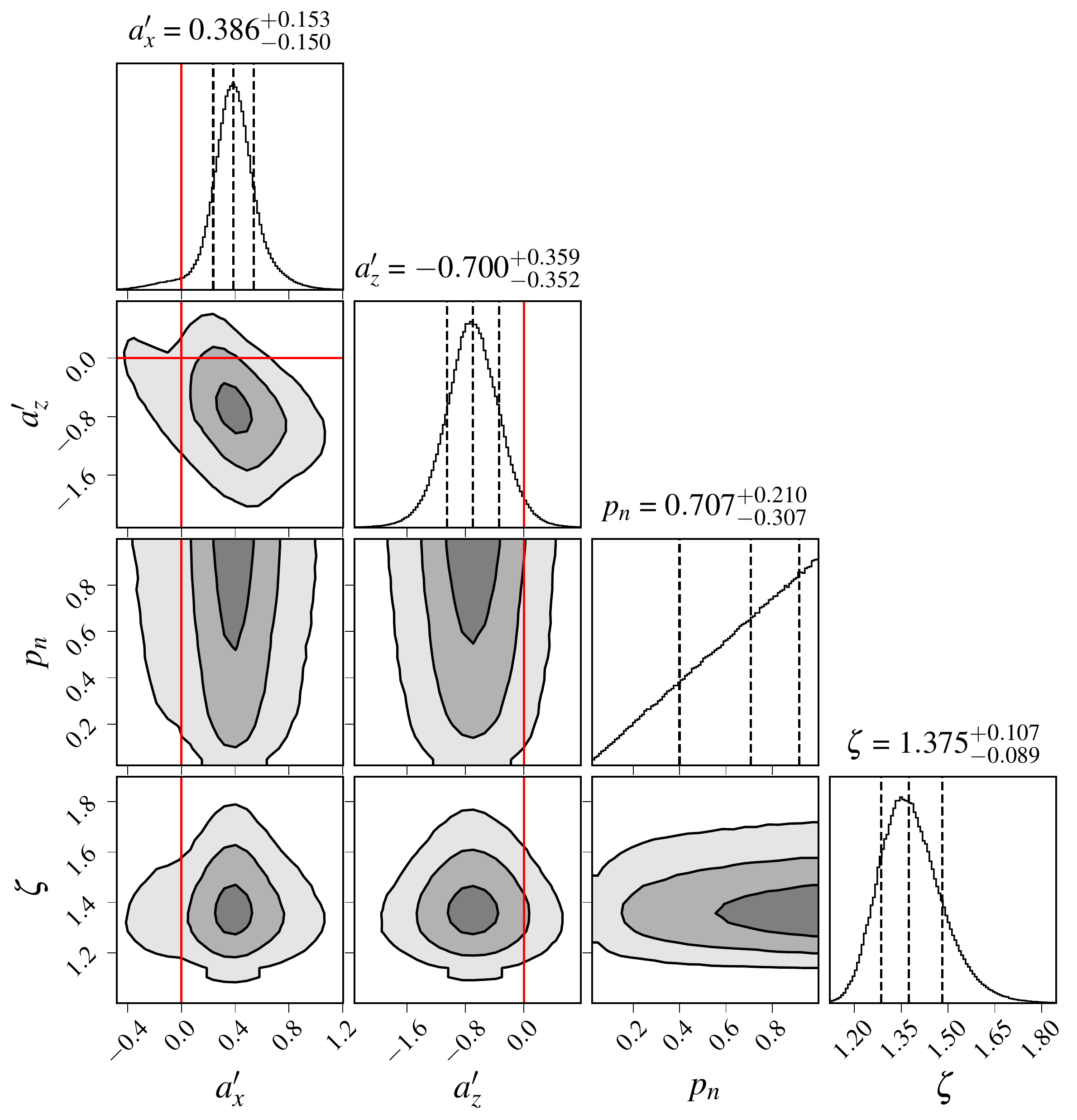}
\caption{Posterior distributions for galactic acceleration gradients in the radial ($a'_x$) and vertical ($a'_z$) directions (see Eq.~\ref{eq:a_gal_model}), the probability ($p_\mathrm{n}$) that a pulsar has an additional acceleration in a random direction, and the spectral index ($\zeta$) for the magnitude of this additional acceleration noise. The acceleration gradients are quoted in units of $\mathrm{[10^{-10} \, m/s^2/kpc]}$. Red vertical and horizontal lines indicate $a'_x = a'_z = 0$, and dashed black lines delineate the 16th, 50th, and 84th percentiles. }
\label{fig:zeta_corner}
\end{figure}

In Fig.~\ref{fig:zeta_corner}, we show the posteriors on these four parameters, which are all detected at moderate to high significance. The MCMC analysis assumes flat priors on all parameters $a_x'$, $a_z'$, $p_n$, and $\zeta$, with $0<p_\mathrm{n}<1$ and $0<\zeta<2$.
The null hypothesis with $a_x' = a_z' = 0$ is excluded at ${2.6}\sigma$ and ${2.0}\sigma$ significance respectively, and as expected, we find that $a'_x > 0$ and $a'_z < 0$. We also find $\zeta = 1.38^{+0.11}_{-0.09}$; such a strong preference for the nuisance acceleration spectral index (and large $p_\mathrm{n}$) indicates the existence of additional noise in the data. The best-fit values $\zeta = 1.38$ and $p_\mathrm{n} = 0.71$ from Fig.~\ref{fig:zeta_corner} are shown in Fig.~\ref{fig:nuisance_model}. Note that the preferred spectral index is significantly larger than unity. In addition, the noise probability is pushing up close to the physically sensible boundary of $p_\mathrm{n}=1$, but since $a_\mathrm{n}^\mathrm{min}$ is so small, this effectively still corresponds to a small probability of \emph{significant} excess noise for all pulsars. {The errors on all derived parameters thus reflect the reality of this excess noise}. 
Future timing data may identify its physical origin, e.g. accretion in these binary systems. 

We summarize the results for the radial and vertical acceleration gradients and the corresponding midplane Galactic disk density (via Eq.~\ref{eq:rho}) in Tab.~\ref{tab:mod_params}. We compare them to the corresponding parameters estimated in Ref.~\cite{Kipper_model} using the orbital arc method~\cite{Kipper_method}, and in Refs.~\cite{holmberg2000local,Creze} using Jeans analyses methods based on \textit{Hipparcos} data. The latter analyses need to assume dynamical equilibrium in the MW, and arrive at values somewhat higher than ours for the disk density, within {3.1} and ${1.8}\sigma$ of our results, respectively. Ref. \cite{Chakrabarti} similarly uses binary pulsars to estimate the parameters of a local acceleration model, but  includes an additional $a'_{\phi}$ term. {Furthermore, Ref. \cite{Chakrabarti} does not include a nuisance parameter; given that we include all pulsars used in this earlier work and find a high probability of noise in the data, the errors on the derived parameters they report may be underestimated. Our radial and vertical acceleration gradient measurements are ${0.3 \sigma}$ and ${2.3} \sigma$ from their values, respectively. The derived densities differ by ${2.0}\sigma$. Due to the our conservative method of handling noise via the acceleration noise parameter, we see only a marginal improvement in the parameter uncertainties despite the use of nearly double the number of sources.}

We introduce a modified chi-squared distribution ($\widetilde{\chi}^2$) to evaluate the model's goodness of fit:
\begin{equation}\label{eq:chi2}
    \widetilde{\chi}^2 = 2\left({\mathrm{erfc}}^{-1}\left[2(1- \mathcal{L}(\delta a_i , \sigma_{a,i}, p_\mathrm{n}, \zeta))\right]\right),
\end{equation}
where $\mathrm{erfc}^{-1}$ is the inverse complementary error function.
This modified distribution is constructed such that $\widetilde{\chi}^2 = n^2$ has the same $p$-value under the likelihood of Eq.~\ref{eq:likelihood} for finite $p_\mathrm{n}$ and $\zeta$ (see Fig.~\ref{fig:nuisance_model}) as $\chi^2 = n^2$ (an $n$-sigma deviation) for the standard chi-square distribution under a Gaussian likelihood. In particular, Eq.~\ref{eq:chi2} approaches the standard $\chi^2$ value in the limit $p_\mathrm{n} \to 0$ or $\zeta\to \infty$. 
In Tab.~\ref{tab:fit}, we show measurements of goodness of fit for the null hypothesis ($a_\mathrm{Gal}\equiv 0$) and Eq.~\ref{eq:a_gal_model} using the best fit parameters shown in Table~\ref{tab:mod_params}. We find that Eq.~\ref{eq:a_gal_model} is a better fit to the acceleration data, both when assuming Gaussian errors ($\chi^2$) and when using the modified distribution $\widetilde{\chi}^2$, which effectively always yields a good fit by construction. 

The results in Tab.~\ref{tab:fit} strongly suggest the presence of additional noise sources which have not previously been taken into account. We handle this excess noise with two nuisance parameters ($p_\mathrm{n}$ and $\zeta$), and show that the direct evidence for acceleration gradients previously suggested in Refs.~\cite{Philips,Chakrabarti,Heflin} survives under this more robust model. Our more comprehensive, non-Gaussian errors establish statistical concordance with the data, as evidenced by the last column of Tab.~\ref{tab:fit}. The physical origin of this excess acceleration noise is unknown, and will be explored in future work. 

\subsection{Modified Catalog}\label{app:mod_catalog}
To validate our selection criteria and analysis methods we have conducted our analysis on the same dataset as in Ref. \cite{Donlon24}, i.e. our current dataset without PSRs B1259-63, J2339-0533, and J0348+0432. 

\begin{figure}[h!]
     \centering
     \includegraphics[width=\columnwidth]{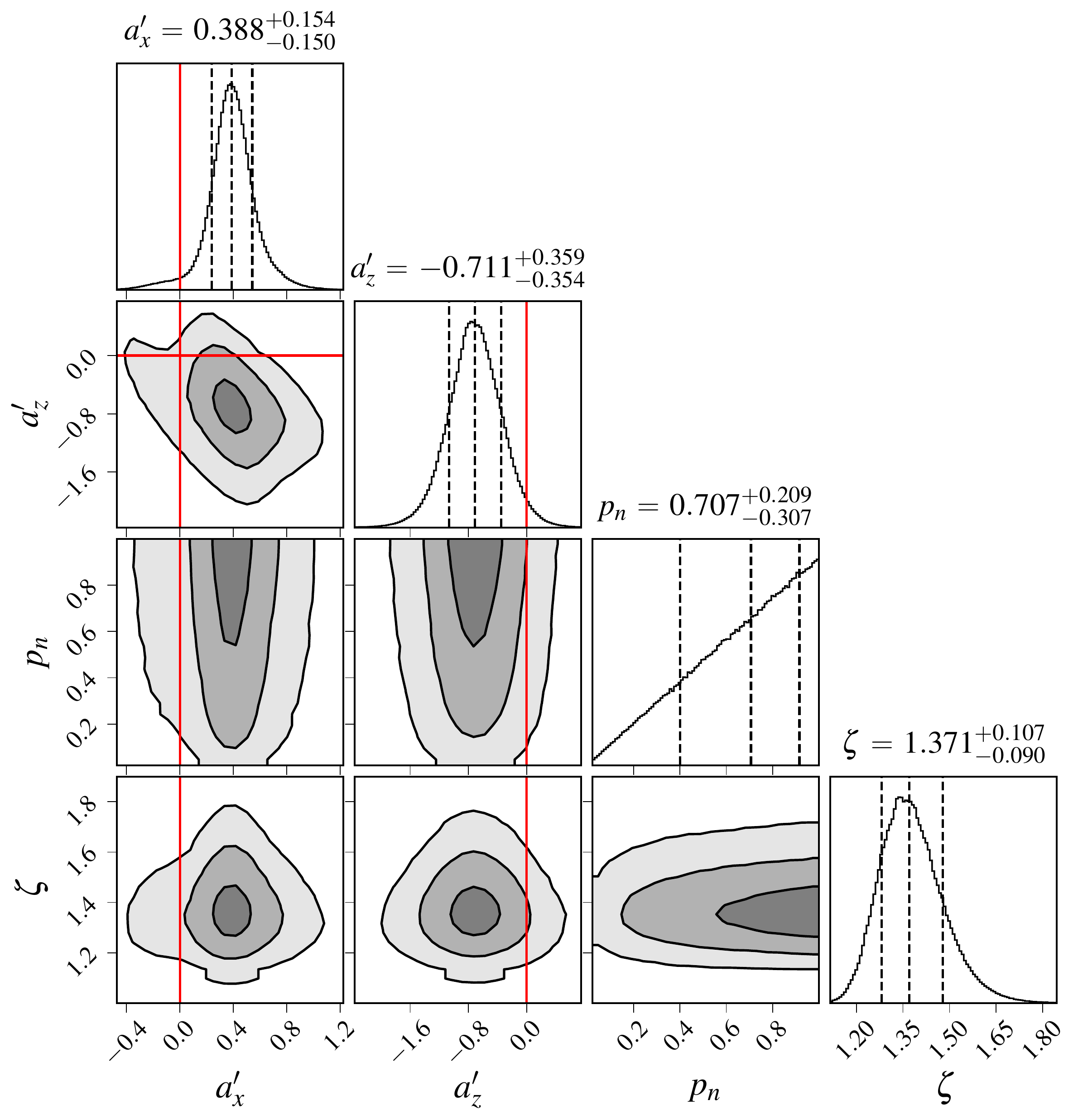}  
    \caption{Posterior distributions using the modified catalog described Sec. \ref{app:mod_catalog}. Shown are the parameters of Eq.~\ref{eq:a_gal_model}, $p_n$ the probability of additional acceleration on a pulsar in a random direction and $\zeta$ the spectral index of this noise. The gradients are in units of [$10^{-10} \mathrm{m}/\mathrm{s}^2/\mathrm{kpc}$]. Red vertical and horizontal lines indicate $a'_x=a'_z=0$ and black dashed lines delineate the 16th, 50th and 84 percentiles.}
     \label{fig:params_2}
 \end{figure}

We find no significant improvement in accuracy (relative to previous indirect measurements) or precision of the derived parameters of our model, as summarized in Tab.~\ref{catalog_comp_tab} and shown in Fig.~\ref{fig:params_2}. We also see no change in either $p_n$ or $\zeta$, indicating that there is still noise in this dataset. This suggests that without our nuisance parameter, the errors on any parameters derived from this catalog will be artificially low. 
This is further supported by the high $\chi^2$ and significantly lower $\widetilde{\chi}^2$ that we find. The same 5 pulsars (B1913+16, 
J1713+0747, J1741+1351, J1933-6211, and J2129-5721) remain significant outliers in this modified catalog, deviating from the model by $35,\ 8.2, \ 3.6,\ 3.2,$ and $10.2 \sigma$ as before. 

We thus find that our more complete catalog does not sacrifice any precision. Furthermore, our robust statistical analysis is indeed necessary even when being more conservative with the selection of pulsars; all possible sources of noise cannot be accounted for in source selection, and this must be accounted for statistically as we have done here.

\begin{table}[b!]
\centering
\renewcommand{\arraystretch}{1.5} 
\begin{tabular}{p{0.55\linewidth} >{\centering\arraybackslash}p{0.2\linewidth} >{\centering\arraybackslash}p{0.2\linewidth}}
\hline 
\hline
    Model & {$\chi^2$} & {$\widetilde{\chi}^2$}\\
    \hline
    No acceleration (null) &  {1181}  &   {41.79}  \\
    Galactic acceleration (Eq.~\ref{eq:a_gal_model}) & {1491} & {38.55} \\
    \hline
    \hline
\end{tabular}
\caption{Goodness of fit for the null hypothesis ($a_{\mathrm{Gal},i} =  0$) and the Galactic acceleration model of Eq.~\ref{eq:a_gal_model}. We report the standard $\chi^2$ values, as well as a modified statistic $\widetilde{\chi}^2$ from Eq.~\ref{eq:chi2} based on the likelihood from Eq.~\ref{eq:likelihood}.}  
\label{tab:fit}
\end{table}

\section{Discussion}\label{sec:discussion}

We present a direct map of Galactic acceleration measurements based on a comprehensive catalog constructed from ATNF. These data can be fit to models of the Galactic potential or matter density~\cite{Chakrabarti}. 
{For the purposes of data validation, we perform parameter estimation for the simple galactic model given in Eq. \ref{eq:a_gal_model}. We} constrain the Galactic midplane density to ${ 0.040_{-0.020}^{+0.020}}M_{\odot} \, \mathrm{pc}^{-3}$ which is within $\sim 3\sigma$ of results from Refs.~\cite{Kipper_model, Creze, Chakrabarti}.
{Using stellar astrometry from \emph{Gaia}, Ref.~\cite{Schutz} reports a local baryon density of $0.0889 \pm 0.0071 \ M_{\odot} \ \mathrm{pc}^{-3}$. Since our estimate of the \emph{total} local density is below this baryonic density, and our uncertainties are still relatively large, we cannot meaningfully constrain the subdominant contribution from the dark matter density, but are hopeful that in future data releases we will be able to do so. }

Our {29} pulsars with 4 fit parameters, reported in Table~\ref{tab:fit}, ``match'' the model of Galactic acceleration given in Eq.~\ref{eq:a_gal_model} with a high Gaussian chi-square value of $\chi^2=1491$. Refs.~\cite{Philips, Chakrabarti} use a similar Gaussian-based method to report their accelerations. 
We find {five} outlier ($\gtrsim 3 \sigma$) acceleration measurements: those from PSRs B1913+16 (${35} \sigma$), J1713+0747 (${8.1} \sigma$), J1741+1351 (${3.6} \sigma$), J1933-6211 (${3.2} \sigma$), and J2129-5721 (${10.3} \sigma$). Of these, PSR B1913+16 lies inside of the thin disk, and the rest are outside of the thin disk, but well inside the thick disk.
The expansion from Eq.~\ref{eq:a_gal_model} thus mildly overestimates the acceleration magnitude for these pulsars, but this mismodeling is not nearly enough to account for the large discrepancies.
{We therefore }reject the Gaussian noise hypothesis, and explore the possibility of an additional spurious noise term contributing to $\dot P_\mathrm{b}^\mathrm{Obs}$. 

We model this with the nuisance parameter ${a}_n$ in Eq.~\ref{eq:chi2}. We find a significant improvement in the subsequent modified $\widetilde{\chi}^2 \approx {39}$ (see Tab.~\ref{tab:fit}) and tight constraints on $\zeta$ (see Fig.~\ref{fig:zeta_corner}). {Moreover, under this modified distribution, there are no $> 3\sigma$ outliers, providing further support for this modified distribution.} This indeed indicates that several of the outliers discussed above are the result of data quality issues (though we eliminated  known systematic effects), intrinsic physical processes such as undetected accretion in the pulsar binary, or rare occurrences of significant acceleration caused by a third body in a wide circumbinary orbit~\cite{Ransom2014}.

The error associated with ${P}_\mathrm{b}$ improves as $T^{-3/2}$ with integration time $T$, so ${\dot{P}}_\mathrm{b}^\text{Obs}$ improves as $T^{-5/2}$~\citep{Damour}. {Furthermore, parallax} errors, used in the computation of $\dot{P}_\mathrm{b}^\text{Kin}$, also improve as $T^{-1/2}$~\citep{Jennings}. Of the {29} pulsar acceleration measurements in our catalog, {16} are limited by ${\dot{P}}_\mathrm{b}^\text{Obs}$ errors, and {7} are dominated by uncertainties on the ${\dot{P}}_\mathrm{b}^\text{Kin}$ determination (primarily parallax error), and {6} are limited by ${\dot{P}}_\mathrm{b}^\text{GW}$.
We are therefore hopeful that our few-sigma measurement of acceleration gradients will turn into a precision probe of Galactic structure with future pulsar timing data releases and correspondingly improved line-of-sight acceleration catalogs.

\acknowledgements
 We thank Sukanya Chakrabarti, William DeRocco, 
Susan Gardner, Matthew Kerr, Mariangela Lisanti, Katelin Schutz, and Oren Slone for helpful conversations and feedback.
 The Center for Computational Astrophysics is a division of the Flatiron Institute in New York City, which is supported by the Simons Foundation. This research was supported in part by the National Science Foundation under Grants No.~NSF PHY-1748958, PHY-2020265, AST-2106552, and PHY-2210551 and by the NASA CT Space Grant Consortium under P-1935, Undergraduate Research Grant.
 This work has made use of data from the European Space Agency (ESA) mission Gaia (https://www.cosmos.esa.int/gaia), processed by the Gaia Data Processing and Analysis Consortium (DPAC, https://www.cosmos.esa.int/web/gaia/dpac/ consortium). Funding for the DPAC has been provided by national institutions, in particular the institutions participating in the Gaia Multilateral Agreement.

\vspace{1cm}
 
\appendix 

\section{Catalog summary}\label{app:catalog}

In Tab.~\ref{tab:bigtab} of this appendix we present the data for each pulsar in our catalog. This includes position and orbital measurements as well as intermediate calculations of $\dot{P}_\mathrm{b}^\mathrm{Kin}$ and final calculations of $a_{\mathrm{Gal}}$. 
Complete data are available in an appended .csv file. 

\begin{figure*}[h!]
     \centering
     \includegraphics[trim={1.2cm 2.8cm .8cm 4.5cm}, clip, width=0.47\textwidth]{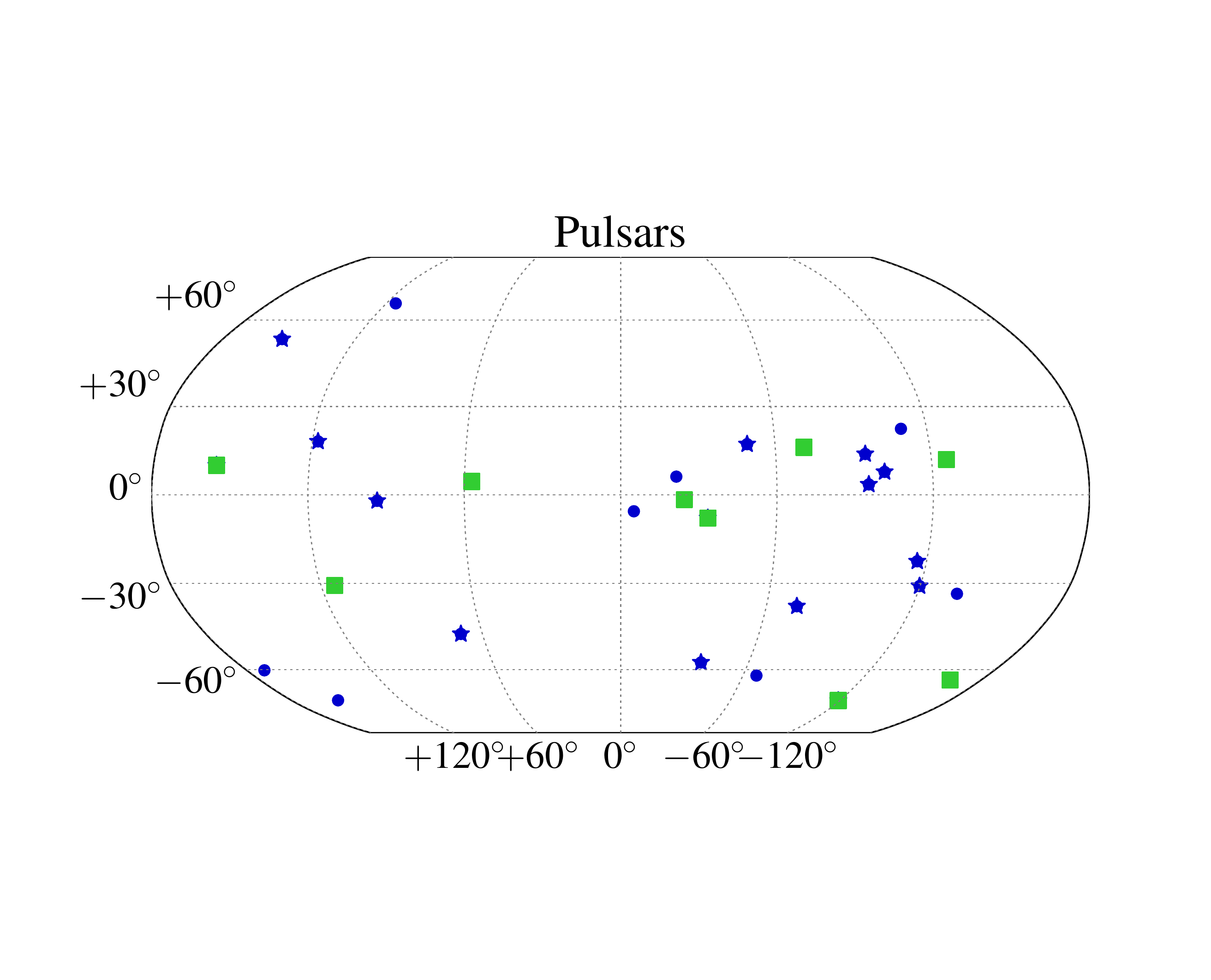}
     \includegraphics[width = 0.46\textwidth]{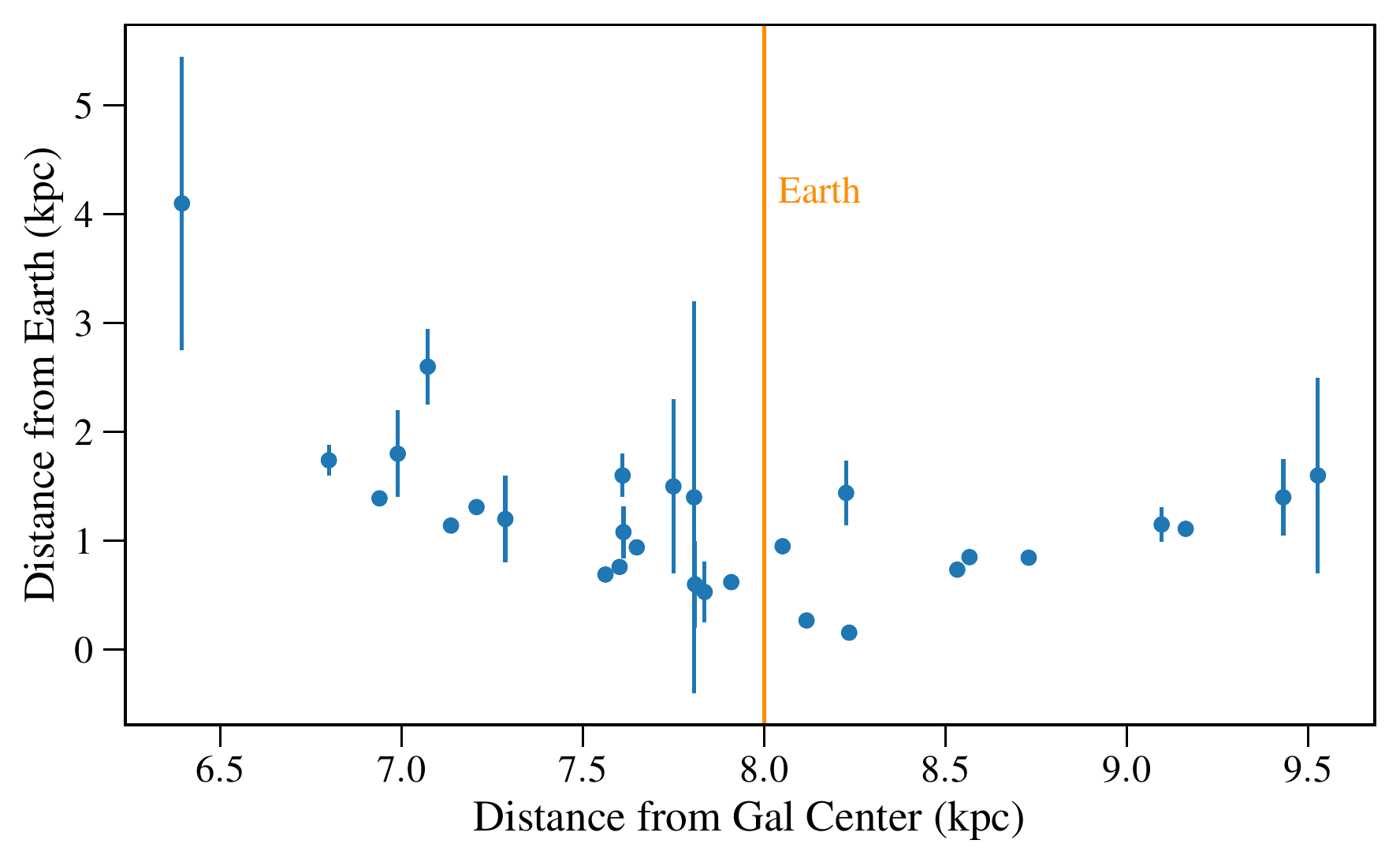}
     \caption{Left: Locations of the {29} pulsars included in this data release {(see Table~\ref{tab:bigtab})}. IPTA and other MSPs are shown in blue, with star and circle markers respectively. Green squares are not MSPs, however they do have measured binary orbital period derivatives. Right: Distances to the {29} pulsars in this data release.}
     \label{fig:skymap}
 \end{figure*}

\begin{figure*}[h!]
 \centering
     \includegraphics[trim={0cm 0cm 0cm .5cm}, width=.49\textwidth]{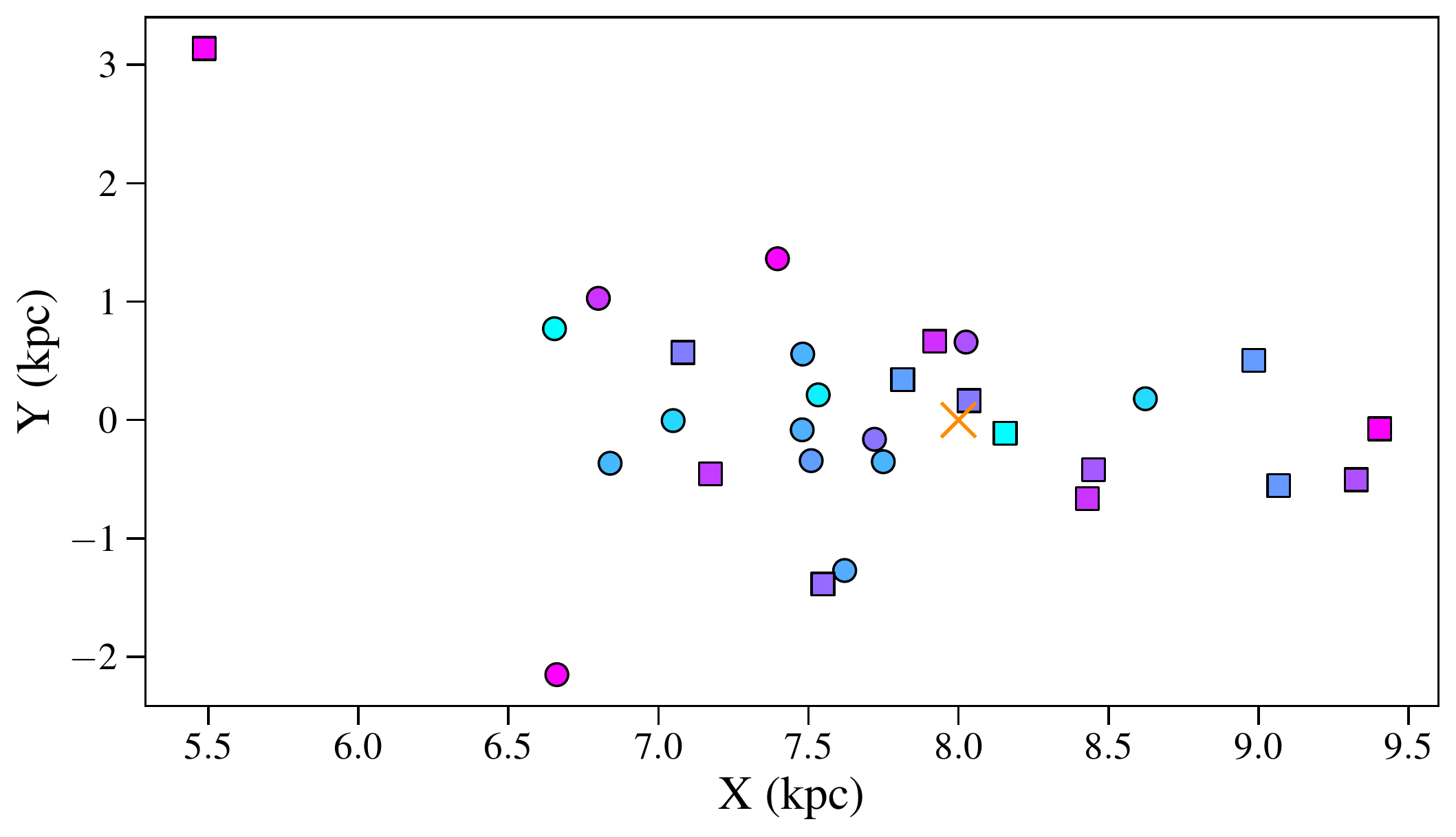}
     \includegraphics[width=.49\textwidth]{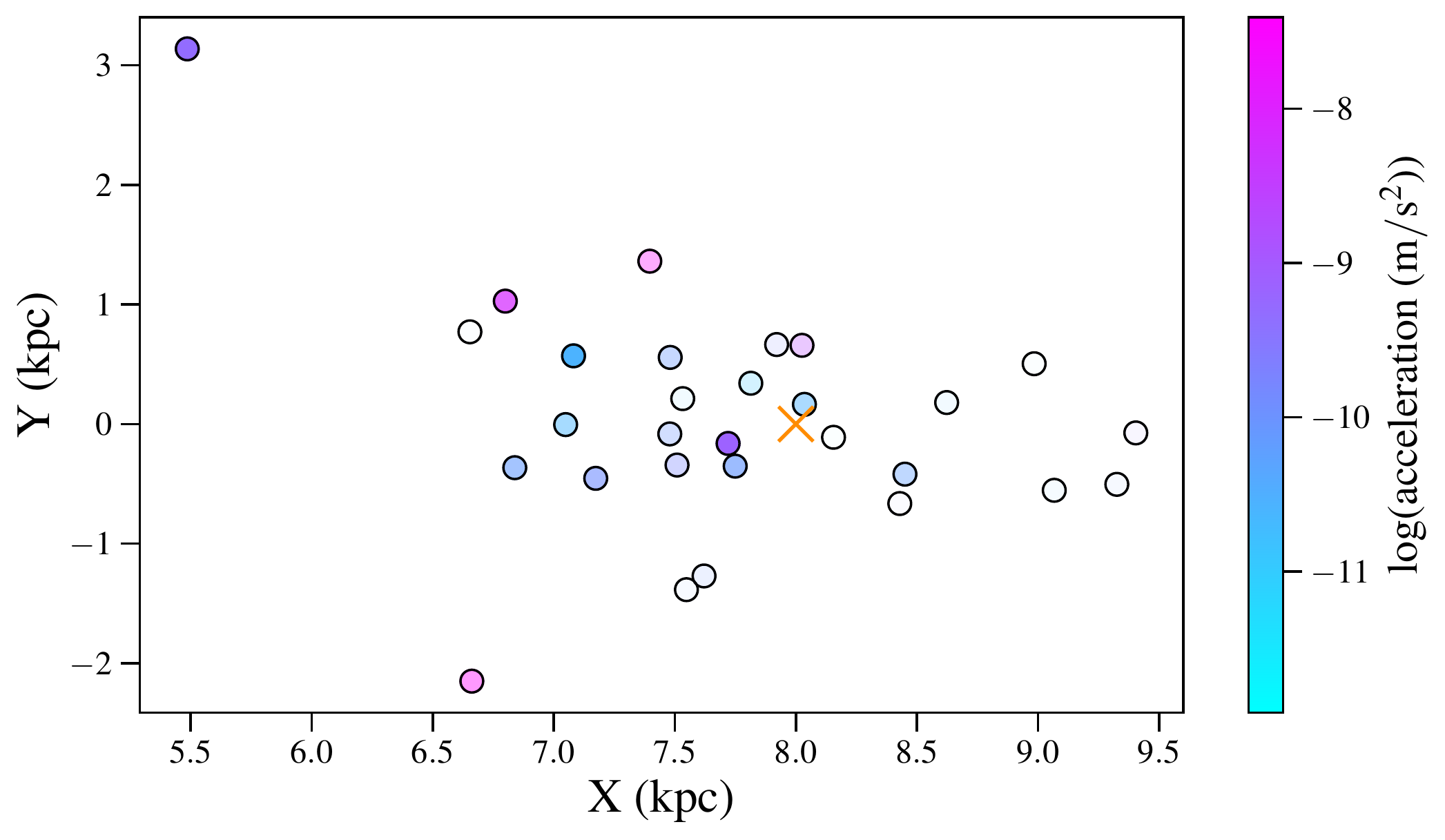}
     \caption{The acceleration data mapped as a function of position in galactocentric coordinates {(see Table~\ref{tab:bigtab})}, from a top-down view of the Milky Way. In both panels, the orange ``X'' is the Earth. The left panel shows the log of the absolute value of the accelerations as a function of position. Square points indicate negative accelerations, while circular ones are positive. The right panel shows the log of the acceleration data with relative uncertainty indicated by opacity; more faint points have higher values of uncertainty relative to calculated acceleration.}
     \label{fig:accmap}
 \end{figure*}

\begin{figure*}[h!]
     \centering
     \includegraphics[width=.55\textwidth]{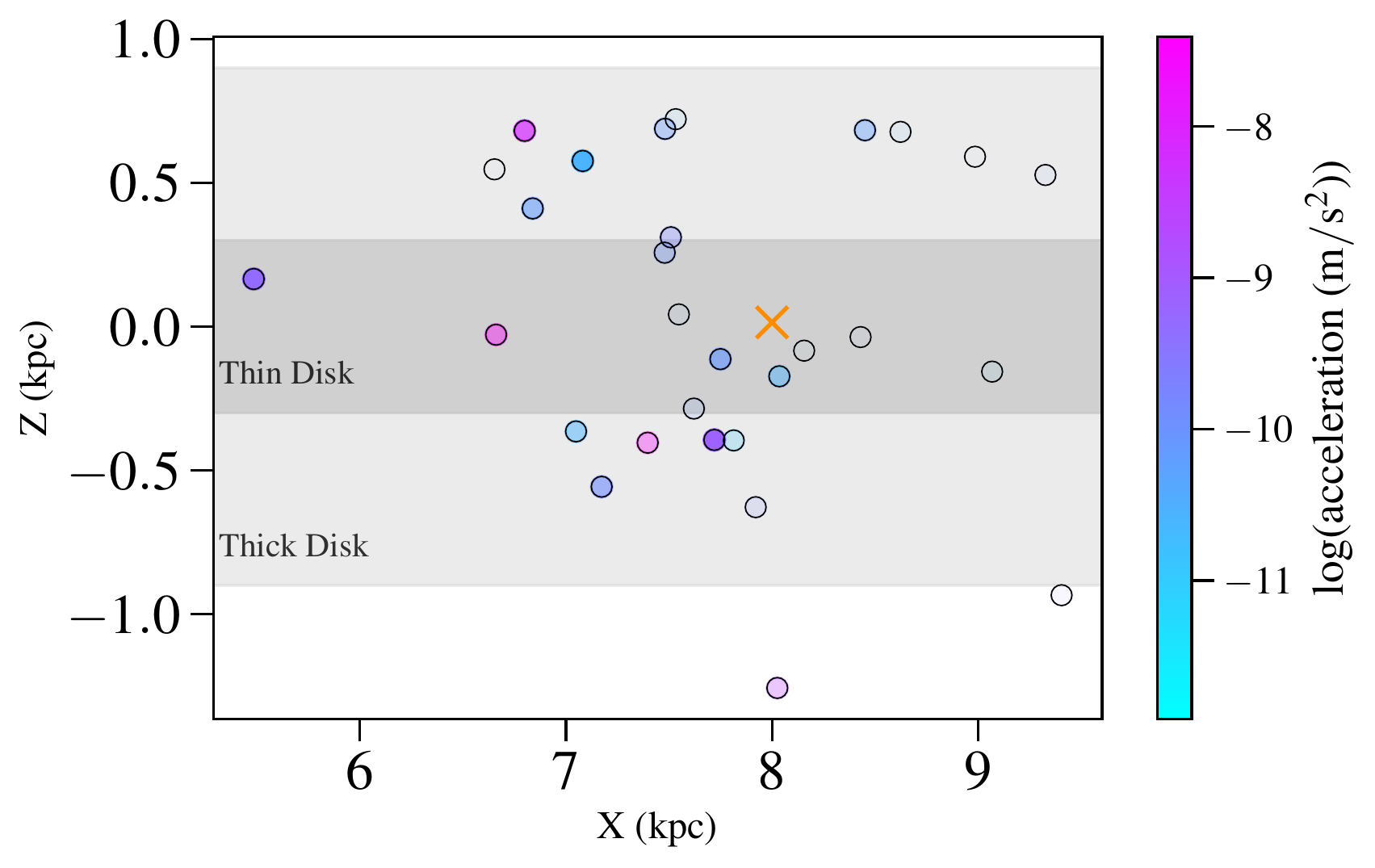}  
    \caption{Acceleration data mapped over an edge on view of the Milky Way in galactocentric coordinates {(see Table~\ref{tab:bigtab})}, where the orange X is the Earth. As in Fig.~\ref{fig:accmap}, more faded points represent acceleration values with higher relative uncertainties. All but two of the pulsars are within the galactic disk, and {11} of the {29} are in the thin disk. }
     \label{fig:zmap}
 \end{figure*}

We also present maps of the pulsars and acceleration data in Figs.~\ref{fig:skymap}, \ref{fig:accmap}, and \ref{fig:zmap}. In addition to sky position, Fig. \ref{fig:skymap} shows the portion of pulsars in our catalog which are monitored by the International Pulsar Timing Array (IPTA) or are millisecond pulsars (MSPs). 
The pulsars are all within 5 kpc of Earth (see Fig. \ref{fig:skymap}). Furthermore, all but two of the pulsars are in the galactic disk, Fig. \ref{fig:zmap}. This distribution is consistent with the assumptions of the radial gradient expansion we use to model galactic acceleration (Eq. \ref{eq:a_gal_model}). 

\begin{table*}[h!]
\begin{center}
\renewcommand{\arraystretch}{1.2}
\caption{\label{tab:bigtab} Data for each pulsar: the galactocentric coordinates $x,y,z$, observed binary period $P_\mathrm{b}$ and its first time derivative $\dot{P}_\mathrm{b}^\mathrm{Obs}$, predicted binary period derivative $\dot{P}_\mathrm{b}^\mathrm{Kin}$ from kinematic effects, our inferred line-of-sight acceleration $a_\mathrm{Gal}$, and the reference used for the distance measurement. A star `*' next to $\mathrm{log}_{10}|\dot{P}_\mathrm{b}^{\mathrm{Obs}}|$ values indicates that $\dot{P}_\mathrm{b}^{\mathrm{Obs}}$ is negative. A dagger `$\dagger$' next to a pulsar's name indicates that $a_\mathrm{Gal}$ is calculated with the approximation $\dot{P}_\mathrm{b}^{\mathrm{GW}} \approx 0$ due to missing mass measurements.}
\setlength{\tabcolsep}{0pt}
\footnotesize
\begin{tabular}{
l  @{\hskip 12pt} 
S[table-format=4.1] @{\hskip 12pt} S[table-format=4.4] @{\hskip 12pt} S[table-format=3.4] @{\hskip 12pt}  
S[table-format=4.4] @{\hskip 12pt} S[table-format=3.2, table-space-text-post = {$^{\dagger}$}] @{\hskip 2pt } >{\text{(}}S[table-format=4.2]<{\text{)}} @{\hskip 12pt} 
S[table-format=4.5] @{\,\( \pm \)\,} S[table-format=3.5] @{\hskip 12pt} 
S[table-format=4.4] @{\,\( \pm \)\,} S[table-format=3.4] @{\hskip 12pt} 
l}
\hline
\hline
Pulsar &  
{$x~[\mathrm{pc}]$}  & {$y~[\mathrm{pc}]$} & {$z~[\mathrm{pc}]$} & 
{$P_\mathrm{b}~[10^{4} \, \mathrm{s} ]$} & \multicolumn{2}{c}{$\mathrm{log}_{10}\big|\dot{P}_\mathrm{b}^{\mathrm{Obs}}\big|$} & 
\multicolumn{2}{c}{$\dot{P}_\mathrm{b}^{\mathrm{Kin}} ~ [10^{-14}]$} & 
\multicolumn{2}{c}{$a_\text{Gal}~ [10^{-10} \mathrm{m \,s^{-2}}]$} & 
Ref.\\
\hline
B1259-63        & 6661.5    & -2150.5   & -27.9   & 10700.00     & -7.85             & -8.15 & 3340      & 450       & 391       & 196       & \cite{MillerJones}\\
B1534+12        & 7532.5    & 212.1    & 721.6    & 3.64    & -12.9* & -15.5 & {5.33}      & 0.37      & 0.106      & 0.313      & \cite{Ding23}\\
B1913+16        & 5486.2    & 3137.1    & 165.9    & 2.79    & {$-$11.6}* & -15.0 & 0.0145    & 0.0086    & -4.74     & 0.15       & \cite{Deller1913}\\
J0348+0432      & {9403.9}     & {$-$74.6}  & {$-$933.8}  & 0.89    & -12.6* & -13.3 & {0.0984}     & {0.060}     & {$-$5.4}     & 18.2      & \cite{Moran} \\
J0437-4715      & {8155.4}    & {$-$111.16}   & {$-$83.4}   & 49.60    & -11.4             & -14.2 & {373.2}       & {2.6}      & {$-$0.122}    & {0.164}    & \cite{Moran} \\
J0613-0200      & 9067.1    & -554.5   & -156.3   & 10.40    &-13.7              & -14.1 & 3.10      & 0.14      & -0.141    & 0.532     & \cite{IPTA}\\
J0737-3039A/B   & {8429.1}    & {$-$665.3}   & {$-$36.1}   & 0.88   & -11.9* & -13.8 & {0.0302}   & {0.0095}     & {$-$1.54}     & 11.3      & \cite{Kramer21} \\
J0740+6620      & 8984.1    & 504.0    & 591.0    & 41.20    & -11.9             & -12.7 & 122       & 17        & 0.14     & 1.91      & \cite{Fonseca} \\
J0751+1807      & 9325.5    & -504.7   & 527.6    & 2.27    & -13.5* & -14.6 & 1.45      & 0.37      & -0.80     & 2.90      & \cite{IPTA} \\
J1012+5307     & {8622.6}    & {179.4}    & {677.4 }   & 5.22    & -13.2             & -14.0 &{ 7.06 }     & 0.12      & {0.224}   & 0.685     & \cite{Moran} \\
J1017-7156      & 7620.6    & -1270.9   & -284.8   & 56.30    & -12.4             & -12.7 & 19.6      & 25.1        & 1.09      & 1.71      & \cite{Rear16}\\
J1022+1001      & {8450.4}    & {$-$419.4}   & {683.2}    & 67.40    & -12.7             & -13.2 & {35.3 }     & {2.29 }     & {$-$0.643}    & {0.320}     & \cite{EPTADR2} \\
J1125-6014      & 7547.9    & -1385.6   & 42.7    & 75.60    & -12.2             & -13.0 & 80.8      & 43.1        & -0.43    & 1.75      & \cite{Rear16} \\
J1455-3330{$^{\dagger}$} & 7509.1   & -343.2   & 310.8   & 658.15    &   -11.3          & -11.7 & 80.7      & 6.5        & 1.72   & 1.01     & \cite{EPTADR2}\\
J1600-3053      & {6838.9}    & {$-$365.4}  & {411.2}   & 124.00    & -12.3             & -13.0 & {20.7}      & {0.62}       & {0.708}     & {0.242}    & \cite{EPTADR2} \\
J1603-7202      & 7749.3    & -352.4   & -112.8   & 5.45    & -12.7             & -13.4 & 4.2      & {2.2}        & 0.815     & {0.252}    & \cite{Rear16}\\
J1614-2230      & 7479.1    & -83.0   & 257.3    & 75.10    & -11.8             & -12.9 & 132       & 10       & 0.984     & 0.653     & \cite{IPTA}\\
J1640+2224 & 7480.8    & 556.8   & 688.1   & 1515.00    & -11.0             & -11.7 & 529       & 118       & 0.833     & 0.448     & \cite{EPTADR2}\\
J1713+0747      & {7081.5}    & {570.0}    & {576.4}    & 586.00    & -12.7             & -13.0 & {73.8}      & {2.3}       & {$-$0.275}    & 0.052    & \cite{PPTADR2} \\
J1738+0333      & {6653.6}    & {770.9}    & {547.3}    & 3.07    & -13.8* & -14.5 & {0.976}     & {0.079}     & {0.062}    & 1.03      & \cite{Ding23}\\
J1741+1351      & 6799.8    & 1027.4    & 681.2  & 5.88   &-11.9              & -12.4 &3.49    & 0.77   & 67.5      & 18.8      & \cite{Nano} \\
J1909-3744     & 7049.0    & -5.0   & -364.3   & 13.20    & -12.3             & -15.2 & 50.1      & 0.4      & 0.246     & 0.101     & \cite{IPTA}\\
J1933-6211     & 7173.3   & -454.6   & -556.6  & 111.00    & -20.4             & -24.7 & 46.9      &  15.7    &  -1.27   &  0.425    & \cite{Geyer}\\
J2043+1711      & 7396.6    & 1361.5    & -403.6  & 0.53   &-13.0              & -13.9 &0.311    & 0.039   & 323      & 194      & \cite{NG15} \\
J2129-5721      & 7720.0    & -162.8   & -393.8   & 57.20    & -11.8             & -13.1 & 14.8      & 9.9       & 7.14      & 0.70     & \cite{PPTADR2} \\
J2145-0750       & 7813.8    & 340.8    & -395.5   & 59.10    & -12.9             & -13.7 & 15.4      & {0.52}      & -0.117    & {0.105}     & \cite{IPTA} \\
J2222-0137      & {8035.2}    & {164.2}    & {$-$172.5  } & 21.10    & -12.6             & -14.1 & {27.9}      & 0.2      & {$-$0.294 } &  0.125     & \cite{Guo}\\
J2234+0611      & 7920.4    & 664.3    & -627.7   & 27.60    & -11.5             & -11.6 & 468       & 20        & -1.71     & 2.72      & \cite{Stovall2234} \\
J2339-0533      & 8025.1    & 657.9    & -1256.5   & 1.67    & -19.7* & -21.7 & 0.722     & 0.167      & 23.2      & 15.7      & \cite{Romani11}\\
\hline
\hline
\end{tabular}
\end{center}
\end{table*}

.

\begin{table*}[h!] 
    \centering
    \renewcommand{\arraystretch}{1.5} 
    \begin{tabular}{l S[table-format=1.6]@{\hspace{0.8em}} S[table-format=1.2]@{\hspace{0.8em}} S[table-format=2.1]@{\hspace{0.8em}} S[table-format=2.1]@{\hspace{0.8em}} S[table-format=2.1]@{\hspace{0.8em}} S[table-format=2.1]@{\hspace{0.8em}} S[table-format=2.1]@{\hspace{0.8em}}}
    \hline 
    \hline
    {Parameter} & {$a'_x$} & {$a_z'$} & {$\rho_\mathrm{d}$} & {$p_n$} & {$\zeta$}& {$\chi^2$} & {$\widetilde{\chi}^2$}\\
    {Units} & {$10^{-10}\,\mathrm{\frac{m}{s^2 \, kpc}}$}  & {$10^{-10}\,\mathrm{\frac{m}{s^2 \, kpc}}$} & {$10^{-2}\,\frac{M_\odot}{\mathrm{pc}^{3}}$}  & {} &{} &{} &{} \\
    \hline
    {Full catalog} & {${0.39_{-0.15}^{+0.15}}$} & {${-0.71^{+0.36}_{-0.35}}$} & {${4.0_{-2.0}^{+2.0}}$} 
    & {${0.71_{-0.3}^{+0.2}}$} &{${1.38_{-0.1}^{+0.1}}$} & {1491} &{38.55}\\
     {Modified catalog} & {${0.39_{-0.15}^{+0.15}}$} & {${-0.70^{+0.36}_{-0.35}}$} & {${4.1_{-2.0}^{+2.0}}$}&{${0.71_{-0.3}^{+0.2}}$} &{${1.37_{-0.1}^{+0.1}}$} & {1486} &{32.34} \\
     \hline
     \hline
    \end{tabular}
    \caption{Best-fit model parameters, distribution shape parameters, and goodness of fit values for the full dataset, and the modified dataset as in Ref.~\cite{Donlon24}. The omission of the three pulsars did not improve our resolution on any of the model parameters, did not decrease the probability of noise in the dataset, and resulted in only a minor improvement in the goodness of fit values.}
    \label{catalog_comp_tab}
\end{table*}

\end{document}